\documentclass[aps,twocolumn,superscriptaddress,preprintnumbers,showpacs]{revtex4}
\usepackage{graphicx}
\usepackage{subfigure}
\usepackage{booktabs}
\usepackage[colorlinks,linkcolor=blue,urlcolor=blue,citecolor=blue,bookmarks,bookmarksnumbered]{hyperref}
\usepackage{enumerate}

\begin{document}
\title{Extracting mass hierarchy information from simple analysis of neutrino mass splitting\footnote{Published in  Mod.\ Phys.\ Lett.\ A {\bf 29}, 1450096 (2014). } }

\newcommand*{\PKU}{School of Physics and State Key Laboratory of Nuclear Physics and Technology, Peking University, Beijing 100871, China}\affiliation{\PKU}
\newcommand*{\CIC}{Collaborative Innovation Center of Quantum Matter, Beijing 100871, China}\affiliation{\CIC}
\newcommand*{\CHEP}{Center for High Energy Physics, Peking University, Beijing 100871, China}\affiliation{\CHEP}
\author{Yijia Zhang}\affiliation{\PKU}
\author{Bo-Qiang Ma}\email{mabq@pku.edu.cn}\affiliation{\PKU}\affiliation{\CIC}\affiliation{\CHEP}

\begin{abstract}
  Based on the independent measurements on neutrino mass splitting $|\Delta m^2_{\mu\mu}|$, $|\Delta m^2_{ee}|$, $\Delta m^2_{21}$, and recent measurements by the T2K Collaboration, we carry out a simple fitting analysis on $\Delta m^2_{32}$ and $\Delta m^2_{31}$ in normal hierarchy and inverted hierarchy respectively, suggesting $\Delta m^2_{32}=(2.46\pm0.07)\times10^{-3}~\mbox{eV}^2$ and $\Delta m^2_{31}=(2.53\pm0.07)\times10^{-3}~\mbox{eV}^2$ in normal hierarchy, or $\Delta m^2_{32}$=$-(2.51\pm0.07)\times10^{-3}~\mbox{eV}^2$ and $\Delta m^2_{31}$=$-(2.44\pm0.07)\times10^{-3}~\mbox{eV}^2$ in inverted hierarchy. The simple analysis indicate that both normal and inverted hierarchy are consistent with current experimental measurements on mass splitting. The p-value for normal hierarchy and that for inverted hierarchy are 62\% and 55\%, respectively. This reveals a slight favor for the normal hierarchy. It is suggested that further measurements on the mass splitting with higher accuracy are necessary to determine the neutrino mass hierarchy.
\end{abstract}


\preprint{DOI:10.1142/S0217732314500965}


\maketitle

\section{\label{sec1}INTRODUCTION}
Through decades of progressive works on neutrino oscillation, neutrinos of three generations with masses and their flavor-mixing properties have been well studied. Increasingly accurate experimental results on the three mixing angles and several mass splitting are continually coming out. Among them, the atmospheric mass splitting $\Delta m^2_{32}$ and the solar mass splitting $\Delta m^2_{21}$ are measured quantities, where $\Delta m^2_{ij}=m^2_i-m^2_j$, and $m_1$, $m_2$, and $m_3$ are masses of the $\nu_1$, $\nu_2$, and $\nu_3$ mass eigenstate neutrinos. It is taken for granted that $|\Delta m^2_{32}|$ is two orders of magnitude larger than $|\Delta m^2_{21}|$~\cite{m21,m32}. In spite of these accurate measurements on the neutrino mass splitting, we still know nothing about whether neutrinos are in normal hierarchy~(NH), i.e., $m_1<m_2<m_3$, or in inverted hierarchy~(IH), i.e., $m_3<m_1<m_2$. Obviously, $\Delta m^2_{31}>\Delta m^2_{32}>0$~($\Delta m^2_{32}<\Delta m^2_{31}<0$) is equivalent to the case of NH~(IH).

The mass hierarchy of neutrinos is a fundamental issue concerning the properties of neutrinos, thus feasibility to determine the mass hierarchy through medium baseline reactor neutrino experiments is explored~\cite{SFGe,Capozzi,Balantekin}. An estimation of $\Delta m^2_{32}$ based on Bernoulli distribution and a Bayesian approach to quantify the confidence level of neutrino mass hierarchy is proposed~\cite{qianxin}. The Bayesian formula for the confidence level of hierarchy is also discussed and derived in Ref.~\cite{Ciuffoli}. Requirements on reactor neutrino experiments and a Fourier analysis method to determine mass hierarchy are studied~\cite{fourier,fourier2,zhanliang}. The JUNO experiment is expected to determine the mass hierarchy at a significance of $4\sigma$ in six years~\cite{caojun}.

In practical experiments, directly measured mass splitting are the absolute values of the effective mass splitting $\Delta m^2_{ee}$ in $\bar{\nu}_e \rightarrow \bar{\nu}_e$~(reactor) mode and $\Delta m^2_{\mu\mu}$ in $\nu_\mu\rightarrow\nu_\mu$ and $\bar{\nu_\mu}\rightarrow\bar{\nu_\mu}$~(accelerator) mode. It is natural trying to determine the neutrino mass hierarchy directly using these two measurements on effective mass splitting $|\Delta m^2_{ee}|$ and $|\Delta m^2_{\mu\mu}|$, and it turns out that the sign of $|\Delta m^2_{ee}|-|\Delta m^2_{\mu\mu}|$ can be used to determine the mass hierarchy~\cite{minus}.

In fact, the effective mass splitting $\Delta m^2_{ee}$ and $\Delta m^2_{\mu\mu}$ are connected to the mass splitting $\Delta m^2_{32}$ and $\Delta m^2_{31}$ by~\cite{caojun,minus}
\begin{eqnarray}
  \Delta m^2_{ee} &=& \eta_e \Delta m^2_{31}+(1-\eta_e)\Delta m^2_{32}\nonumber\\
  && + {\cal O}(\Delta m^2_{ee}\cdot\Delta^2),\label{equee}\\
  \Delta m^2_{\mu\mu} &=& \eta_\mu \Delta m^2_{31}+(1-\eta_\mu)\Delta m^2_{32}\nonumber\\
  && +{\cal O}(\Delta m^2_{\mu\mu}\cdot\Delta^2), \label{equmm}
\end{eqnarray}
where
\begin{eqnarray}
  \eta_\alpha &=& \frac{|\rm U_{\alpha1}|^2}{|\rm U_{\alpha1}|^2+|\rm U_{\alpha2}|^2}+{\cal O}(\Delta^2),\label{eta}\\
  \Delta &=& \frac{\Delta m^2_{21}L}{4E}=\frac{1.267\times\Delta m^2_{21}[\mbox{eV}^2]L[\mbox{km}]}{E[\mbox{GeV}]},\label{LE}
\end{eqnarray}
with $L$ in Eq.~(\ref{LE}) being the distance traveled by the neutrino and $E$ being its energy. In the MINOS experiment, $L/E\sim250~\mbox{km/GeV}$~\cite{mmm}, and in the Daya Bay experiment, $L/E\sim500~\mbox{km/GeV}$~\cite{mee}. For these $L/E$ values, $\Delta^2\sim0.002$. Detailed calculations show that the higher-order terms of Eqs.~(\ref{equee}), (\ref{equmm}) and (\ref{eta}) introduce a relative error about $0.06\%$, which is small enough for us to neglect these higher-order terms. From Eqs.~(\ref{equee}) and (\ref{equmm}), we have
\begin{equation}
  |\Delta m^2_{ee}|-|\Delta m^2_{\mu\mu}|=(\eta_e-\eta_\mu)(|\Delta m^2_{31}|-|\Delta m^2_{32}|) \label{minus}.
\end{equation}
Here, $\eta_\alpha$ is derived from the absolute values of the elements in the Pontecorvo-Maki-Nakagawa-Sakata~(PMNS) mixing matrix~\cite{PMNS}. Referring to the PMNS matrix in the standard parametrization~\cite{CK}, these absolute values are then completely determined by the three neutrino mixing angles $\theta_{12}$, $\theta_{23}$, $\theta_{13}$~\cite{mixing} and one unknown {\it CP} phase $\delta$
\begin{eqnarray}
  \mbox{NH:}\qquad\sin^2\theta_{12}&=&0.307^{+0.018}_{-0.016},\nonumber\\
  \sin^2\theta_{23}&=&0.386^{+0.024}_{-0.021},\label{angle1}\\
  \sin^2\theta_{13}&=&0.0241\pm0.0025.\nonumber
\end{eqnarray}
\begin{eqnarray}
  \mbox{IH:}\qquad\sin^2\theta_{12}&=&0.307^{+0.018}_{-0.016},\nonumber\\
  \sin^2\theta_{23}&=&0.392^{+0.039}_{-0.022},\label{angle2}\\
  \sin^2\theta_{13}&=&0.0244^{+0.0023}_{-0.0025}.\nonumber
\end{eqnarray}
Because there are no measurements on the {\it CP} phase $\delta$ yet, in the following calculations we simply set $\cos\delta=0\pm1$. The effect of $\delta$ on the fitting results is analyzed in Sec.~\ref{sec4}.

From Eqs.~(\ref{eta}) and (\ref{angle1}), we arrive at
\begin{eqnarray}
  \eta_e &=& 0.693^{+0.016}_{-0.018}, \\
  \eta_\mu &=& 0.326^{+0.100}_{-0.126}, \\
  \eta_e-\eta_\mu &=& 0.367^{+0.129}_{-0.105}. \label{minuseta}
\end{eqnarray}
That is, we have $\eta_e-\eta_\mu>0$ at a significance of $3\sigma$. Consequently, the sign of $|\Delta m^2_{ee}|-|\Delta m^2_{\mu\mu}|$ is the same as the sign of $|\Delta m^2_{31}|-|\Delta m^2_{32}|$. Thus, if the sign of $|\Delta m^2_{ee}|-|\Delta m^2_{\mu\mu}|$ is plus~(minus), neutrinos are in NH~(IH).

Up to now, there is only one available measurement on the effective mass splitting $|\Delta m^2_{ee}|$ by the Daya Bay Collaboration~\cite{mee}. Experimental results of $|\Delta m^2_{\mu\mu}|$ measured by the MINOS Collaboration are continually updated~\cite{m32,mmm}. It deserves to mention that the MINOS result~(in 2011) $|\Delta m^2|= 2.32^{+0.12}_{-0.08}\times10^{-3}~\mbox{eV}^2$~\cite{m32} is the one adopted by the Particle Data Group as the recommended value for $\Delta m^2_{32}$~\cite{PDG}. Accurately speaking, this $|\Delta m^2|$ measured by the MINOS Collaboration is actually the effective mass splitting $|\Delta m^2_{\mu\mu}|$, different from $\Delta m^2_{32}$ by about $2\%$. The MINOS result in Ref.~\cite{mmm} is an update to their former result in Ref.~\cite{m32}~(in 2011). In this article, we use the latest results~\cite{mee,mmm}
\begin{eqnarray}
  |\Delta m^2_{ee}| &=& 2.59^{+0.19}_{-0.20}\times10^{-3}~\mbox{eV}^2, \label{mee}\\
  |\Delta m^2_{\mu\mu}| &=& 2.41^{+0.09}_{-0.10}\times10^{-3}~\mbox{eV}^2. \label{mmm}
\end{eqnarray}
From Eqs.~(\ref{mee}) and (\ref{mmm}), $|\Delta m^2_{\mu\mu}|$ is inside the $1\sigma$ error range of $|\Delta m^2_{ee}|$. That means that we cannot determine the sign of $|\Delta m^2_{ee}|-|\Delta m^2_{\mu\mu}|$ even at a significance of $1\sigma$. Combined with Eqs.~(\ref{minus}) and (\ref{minuseta}), the sign of $|\Delta m^2_{31}|-|\Delta m^2_{32}|$ cannot be determined, either. Thus, we cannot draw a conclusion on the mass hierarchy from the analysis above.

\begin{figure} 
  \centering
  \includegraphics[width=8.6cm]{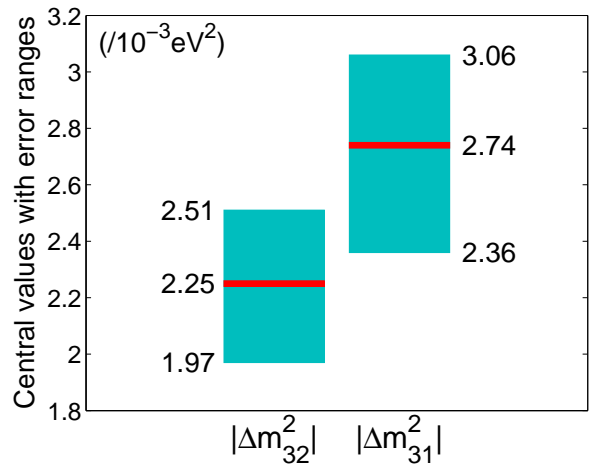}\\
  \caption{$|\Delta m^2_{32}|$ and $|\Delta m^2_{31}|$ calculated from $|\Delta m^2_{ee}|$ and $|\Delta m^2_{\mu\mu}|$. Since the sign of $|\Delta m^2_{32}|-|\Delta m^2_{31}|$ cannot be determined by these calculations, no meaningful conclusions on neutrino mass hierarchy can be derived from these results.}\label{figem}
\end{figure}

To be more accurate, through calculations we arrive at the results
\begin{eqnarray} 
  |\Delta m^2_{32}| &=& 2.25^{+0.26}_{-0.28}\times10^{-3}~\mbox{eV}^2, \\
  |\Delta m^2_{31}| &=& 2.74^{+0.32}_{-0.38}\times10^{-3}~\mbox{eV}^2, \\
  |\Delta m^2_{31}|-|\Delta m^2_{32}| &=& (4.9\pm6.1)\times10^{-4}~\mbox{eV}^2. \label{minusexp}
\end{eqnarray} 
These results are also displayed in Fig.~\ref{figem}.

From Eq.~(\ref{minusexp}) or Fig.~\ref{figem}, we can draw no conclusions on the sign of $|\Delta m^2_{31}|-|\Delta m^2_{32}|$. That is, either $|\Delta m^2_{31}|>|\Delta m^2_{32}|$ or $|\Delta m^2_{31}|<|\Delta m^2_{32}|$ fits the data. By the normal cumulative distribution function, this result in Eq.~(\ref{minusexp}) can be interpreted as a favor for NH with a p-value to be 58\%.

In the calculations and analysis above, we have used mixing angles in Eq.~(\ref{angle1}) of NH since the small differences between Eqs.~(\ref{angle1}) and (\ref{angle2}) cannot alter the conclusions. In the rest of this article, Eqs.~(\ref{angle1}) and (\ref{angle2}) will be used for calculations in NH and IH, respectively.

\section{\label{sec2}SIMPLE FITTING ANALYSIS OF MASS SPLITTING}
Through our efforts to determine neutrino mass hierarchy in Sec.~\ref{sec1}, we realize that the two effective mass splitting are by far insufficient to draw a conclusion, and that more mass splitting measurements are necessary. Fortunately, there are also measurements on the solar neutrino mass splitting $\Delta m^2_{21}$. In this article, we use the value of $\Delta m^2_{21}$ measured by the KamLAND Collaboration~\cite{m21} and recommended by the Particle Data Group~\cite{PDG}
\begin{equation}
  \Delta m^2_{21}=(7.50^{+0.19}_{-0.20})\times10^{-5}~\mbox{eV}^2, \label{m21}
\end{equation}
where $\Delta m^2_{21}$ is connected with other mass splitting by
\begin{equation}\label{equ}
  \Delta m^2_{31}-\Delta m^2_{32}=\Delta m^2_{21}.
\end{equation}

Since there are no conclusions from the analysis in Sec.~\ref{sec1}, we consider it useful to carry out a simple analysis for the two mass splitting $\Delta m^2_{32}$ and $\Delta m^2_{31}$ by adopting the three constraints Eqs.~(\ref{equee}), (\ref{equmm}), and (\ref{equ}).

According to the conventional weighted $\chi^2$ test, we regard $\Delta m^2_{32}$ and $\Delta m^2_{31}$ as free parameters, and express the corresponding $\Delta m^{2(\rm fit)}_{ee}$, $\Delta m^{2(\rm fit)}_{\mu\mu}$, and $\Delta m^{2(\rm fit)}_{21}$ by the two parameters. Then, we minimize
\begin{eqnarray}
  \chi^2 &=& \frac{1}{\sigma^2_{ee}}(\Delta m^{2(\rm fit)}_{ee}-\Delta m^{2(\rm exp)}_{ee})^2 \nonumber\\
  &&+\frac{1}{\sigma^2_{\mu\mu}}(\Delta m^{2(\rm fit)}_{\mu\mu}-\Delta m^{2(\rm exp)}_{\mu\mu})^2 \nonumber\\
  &&+\frac{1}{\sigma^2_{21}}(\Delta m^{2(\rm fit)}_{21}-\Delta m^{2(\rm exp)}_{21})^2, \label{chi2}
\end{eqnarray}
where $\Delta m^{2(\rm exp)}_{ee}$, $\Delta m^{2(\rm exp)}_{\mu\mu}$, $\Delta m^{2(\rm exp)}_{21}$ are the corresponding experimentally observed values, and $\sigma_{ee}$, $\sigma_{\mu\mu}$, $\sigma_{21}$ are their experimental errors, respectively. In this fitting, the degree of freedom~(DoF) is $3-2=1$.

\begin{figure} 
  \centering
  \includegraphics[width=8.6cm]{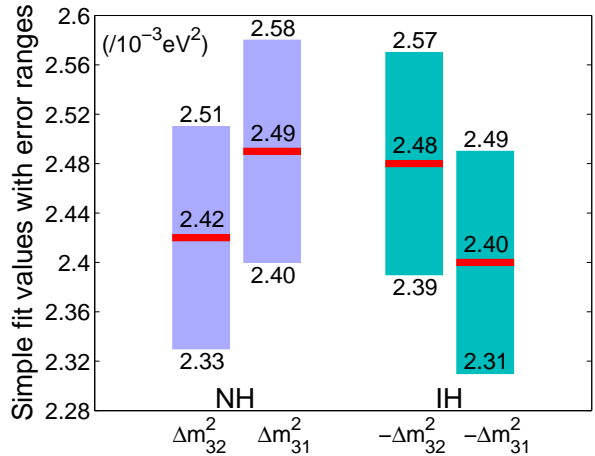}\\
  \caption{Simple fitting results for $\Delta m^2_{32}$ and $\Delta m^2_{31}$ in normal hierarchy~(NH) and inverted hierarchy~(IH), respectively, using Eqs.~(\ref{mee}), (\ref{mmm}), and (\ref{m21}) as constraints.}\label{figfit}
\end{figure}

After detailed calculations, we arrive at the results in Table~\ref{tabfit}, where all the errors from mixing angles, {\it CP} phase, and mass splitting are taken into account.
\begin{table} 
  \caption{Simple fitting for mass splitting $\Delta m^2_{32}$ and $\Delta m^2_{31}$ using Eqs.~(\ref{mee}), (\ref{mmm}), and (\ref{m21}) as constraints. The last row represents the corresponding 2-tailed p-values according to $\chi^2/\mbox{DoF}$~(Degree of Freedom), where a slight preference for normal hierarchy is disclosed.}\label{tabfit}
  \begin{ruledtabular}
  \begin{tabular}{ccc}
    \toprule
    & Fit in normal hierarchy & Fit in inverted hierarchy \\
    \hline
    $\Delta m^2_{32}$ & $(2.42\pm0.09)\times10^{-3}~\mbox{eV}^2$ & $-(2.48\pm0.09)\times10^{-3}~\mbox{eV}^2$ \\
    $\Delta m^2_{31}$ & $(2.49\pm0.09)\times10^{-3}~\mbox{eV}^2$ & $-(2.40\pm0.09)\times10^{-3}~\mbox{eV}^2$ \\
    $\chi^2/\mbox{DoF}$ & $0.46/1$ & $0.86/1$ \\
    p-value & $50\%$ & $35\%$ \\
    \bottomrule
  \end{tabular}
  \end{ruledtabular}
\end{table} 
The last row of Table~\ref{tabfit} are the 2-tailed p-values derived by $\chi^2$ cumulative distribution function, representing the corresponding confidence levels. The simple fitting values of $\Delta m^2_{32}$ and $\Delta m^2_{31}$ are also illustrated in Fig.~\ref{figfit}.

In Table~\ref{tabfit}, the value of $\chi^2/\mbox{DoF}$ in NH, $0.46/1$, is smaller than that in IH, $0.86/1$. Correspondingly, NH is suggested with a p-value of $50\%$, larger than the p-value of IH, i.e., $35\%$. Therefore, both normal and inverted hierarchy are consistent with current experiments, and from the analysis we figure out a slight favor for NH than IH.

\section{\label{sec3}FITTING COMBINED WITH RECENT T2K RESULTS}
Recently, the T2K Collaboration has announced their best-fit mass-squared splitting measured from muon neutrino disappearance experiment~\cite{T2K}:
\begin{eqnarray}
  && \mbox{Assuming NH:  }\Delta m^2_{32}=(2.51\pm0.10)\times10^{-3}~\mbox{eV}^2, \label{m32}\\
  && \mbox{Assuming IH:  }\Delta m^2_{31}=-(2.48\pm0.10)\times10^{-3}~\mbox{eV}^2. \label{m31}
\end{eqnarray}
We can simply add these constraints to our fitting by slightly changing the $\chi^2$ functions. For NH, we use
\begin{eqnarray}
  \chi^2_{\rm NH} &=& \frac{1}{\sigma^2_{ee}}(\Delta m^{2(\rm fit)}_{ee}-\Delta m^{2(\rm exp)}_{ee})^2 \nonumber\\
  &&+\frac{1}{\sigma^2_{\mu\mu}}(\Delta m^{2(\rm fit)}_{\mu\mu}-\Delta m^{2(\rm exp)}_{\mu\mu})^2 \nonumber\\
  &&+\frac{1}{\sigma^2_{21}}(\Delta m^{2(\rm fit)}_{21}-\Delta m^{2(\rm exp)}_{21})^2 \nonumber\\
  &&+\frac{1}{\sigma^2_{32}}(\Delta m^{2(\rm fit)}_{32}-\Delta m^{2(\rm exp)}_{32})^2. \label{chi2NH}
\end{eqnarray}
For IH, we use
\begin{eqnarray}
  \chi^2_{\rm IH} &=& \frac{1}{\sigma^2_{ee}}(\Delta m^{2(\rm fit)}_{ee}-\Delta m^{2(\rm exp)}_{ee})^2 \nonumber\\
  &&+\frac{1}{\sigma^2_{\mu\mu}}(\Delta m^{2(\rm fit)}_{\mu\mu}-\Delta m^{2(\rm exp)}_{\mu\mu})^2 \nonumber\\
  &&+\frac{1}{\sigma^2_{21}}(\Delta m^{2(\rm fit)}_{21}-\Delta m^{2(\rm exp)}_{21})^2 \nonumber\\
  &&+\frac{1}{\sigma^2_{31}}(\Delta m^{2(\rm fit)}_{31}-\Delta m^{2(\rm exp)}_{31})^2. \label{chi2IH}
\end{eqnarray}

\begin{figure} 
  \centering
  \includegraphics[width=8.6cm]{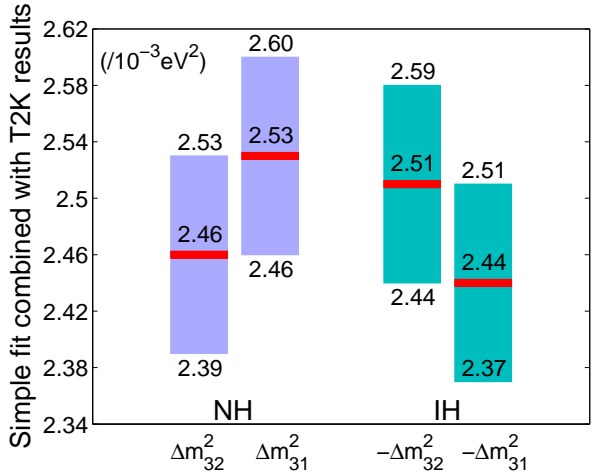}\\
  \caption{Fitting results for $\Delta m^2_{32}$ and $\Delta m^2_{31}$ in normal hierarchy~(NH) and inverted hierarchy~(IH) using Eqs.~(\ref{mee}), (\ref{mmm}), (\ref{m21}), and the T2K measurements Eq.~(\ref{m32}) in NH~(or (\ref{m31}) in IH) as constraints.}\label{figt2k}
\end{figure}

After calculations, we list the fitting results in Table~\ref{tabfit2}, and we also draw the central fitting values and the error ranges in Fig.~\ref{figt2k}. In this fitting analysis combined with constraints from the T2K experiment, the p-value of NH increases to 62\%, and the p-value of IH increases to 55\%. Though not able for us to draw a $1\sigma$ level conclusion, there remains a slight favor for NH.
\begin{table} 
  \caption{Simple fitting for mass splitting $\Delta m^2_{32}$ and $\Delta m^2_{31}$ using Eqs.~(\ref{mee}), (\ref{mmm}), (\ref{m21}), and (\ref{m32}) in NH~(or (\ref{m31}) in IH) as constraints. The corresponding 2-tailed p-values increase from that in Table~\ref{tabfit}. Here the slight preference for normal hierarchy remains.}\label{tabfit2}
  \begin{ruledtabular}
  \begin{tabular}{ccc}
    \toprule
    & Fit in normal hierarchy & Fit in inverted hierarchy \\
    \hline
    $\Delta m^2_{32}$ & $(2.46\pm0.07)\times10^{-3}~\mbox{eV}^2$ & $-(2.51\pm0.07)\times10^{-3}~\mbox{eV}^2$ \\
    $\Delta m^2_{31}$ & $(2.53\pm0.07)\times10^{-3}~\mbox{eV}^2$ & $-(2.44\pm0.07)\times10^{-3}~\mbox{eV}^2$ \\
    $\chi^2/\mbox{DoF}$ & $0.96/2$ & $1.21/2$ \\
    p-value & $62\%$ & $55\%$ \\
    \bottomrule
  \end{tabular}
  \end{ruledtabular}
\end{table}

\section{\label{sec4}DISCUSSIONS AND CONCLUSIONS}
\begin{figure} 
  \centering
  \includegraphics[width=8.6cm]{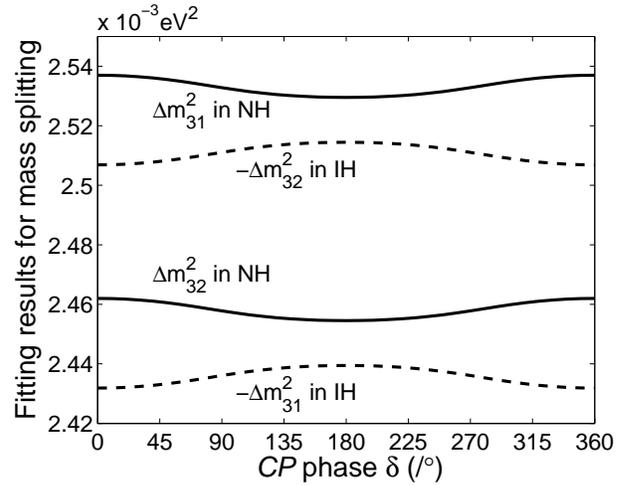}\\
  \caption{Fitting results for $\Delta m^2_{32}$ and $\Delta m^2_{31}$ in normal hierarchy~(NH) and inverted hierarchy~(IH) in different setting of the {\it CP} phase. The solid lines are for NH, and the dotted lines are for IH.}\label{figmass}
\end{figure}

\begin{figure} 
  \centering
  \includegraphics[width=8.6cm]{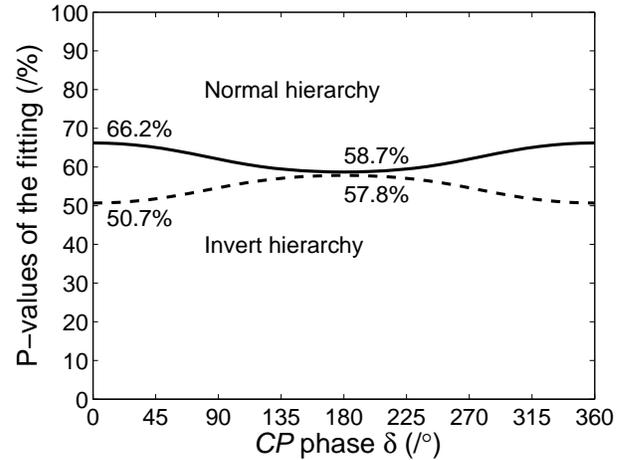}\\
  \caption{P-values for the fitting in normal hierarchy~(NH) and inverted hierarchy~(IH) in different setting of the {\it CP} phase. The solid line is for NH, and the dotted line is for IH.}\label{figprob}
\end{figure}

In the fitting analysis in Sec.~\ref{sec2} and Sec.~\ref{sec3}, we have simply set $\cos\delta=0\pm1$. To see how the {\it CP} phase $\delta$ affects the fitting results, we carried out the fitting for different $\delta$. The results illustrated in Fig.~\ref{figmass} and Fig.~\ref{figprob} show that the slight favor for NH is undisturbed by different {\it CP} phase setting.

Since neutrinos must be in either NH or IH in the three-generation neutrino framework, we can try to combine the two fitting results in Table~\ref{tabfit2} to construct a relative preference for NH and IH from the Bayesian point of view~\cite{qianxin}. The spirit of this Bayesian approach is adjusting our estimation of the reality to the information we gathered. In the following discussion, we denote the collected experimental data by $x$. Consequently, $P(\mbox{NH}|x)$ and $P(\mbox{IH}|x)$ stand for our subjective preference for NH and the preference for IH based on the data, and there must be $P(\mbox{NH}|x)+P(\mbox{IH}|x)=1$.

According to Bayes' theorem, there are
\begin{eqnarray}
  P(\mbox{NH}|x) &=& \frac{P(x|\mbox{NH})\cdot P(\mbox{NH})}{P(x)} \nonumber\\
  &=& \frac{P(x|\mbox{NH})\cdot P(\mbox{NH})}{P(x|\mbox{NH})\cdot P(\mbox{NH})+P(x|\mbox{IH})\cdot P(\mbox{IH})} \nonumber\\
  &=& \frac{P(x|\mbox{NH})}{P(x|\mbox{NH})+P(x|\mbox{IH})} \label{Bayesian},
\end{eqnarray}
where $P(\mbox{NH})$ and $P(\mbox{IH})$ stand for our preferences for NH and IH before we know the data, and we have used simply $P(\mbox{NH})=P(\mbox{IH})=50\%$. From our results in Table~\ref{tabfit2}, we have $P(x|\mbox{NH})=62\%$ and $P(x|\mbox{IH})=55\%$. Together with Eq.~(\ref{Bayesian}), these finally lead to our relative preferences for NH and IH in the Bayesian viewpoint:
\begin{eqnarray}
  P(\mbox{NH}|x) &=& 53\%, \label{bayesp}\\
  P(\mbox{IH}|x) &=& 47\%.
\end{eqnarray}
Thus, the preference ratio of normal vs. inverted mass hierarchy is 53\% vs. 47\% in the Bayesian approach.

The results in Table~\ref{tabfit}, Table~\ref{tabfit2}, Eq.~(\ref{bayesp}), and Fig.~\ref{figprob} indicate a slight preference for NH. Nonetheless, we cannot draw a stronger conclusion because of the low accuracy of the mass splitting measurements. When there are more accurate experimental values for the mass splitting in the future, our simple fitting method will be more useful to settle the neutrino mass hierarchy problem. In addition, our simple fitting method is able to figure out possible disagreement among different measurements on the mass splitting $\Delta m^2_{ee}$, $\Delta m^2_{\mu\mu}$, $\Delta m^2_{21}$, $\Delta m^2_{32}$, and $\Delta m^2_{31}$. Possible conflicts appearing from the fitting analysis could reveal new physics beyond the three-generation neutrino framework.

In conclusion, we suggest an analysis to determine the neutrino mass hierarchy using the available measurements on the mass splitting $|\Delta m^2_{ee}|$ in Eq.~(\ref{mee}), $|\Delta m^2_{\mu\mu}|$ in Eq.~(\ref{mmm}), $\Delta m^2_{21}$ in Eq.~(\ref{m21}), $\Delta m^2_{32}$ in Eq.~(\ref{m32}), and $\Delta m^2_{31}$ in Eq.~(\ref{m31}). We carry out a simple fitting analysis for the mass splitting $\Delta m^2_{32}$ and $\Delta m^2_{31}$~(Table~\ref{tabfit}, Table~\ref{tabfit2}, Fig.~\ref{figfit}, and Fig.~\ref{figt2k}), suggesting $\Delta m^2_{32}=(2.46\pm0.07)\times10^{-3}~\mbox{eV}^2$ and $\Delta m^2_{31}=(2.53\pm0.07)\times10^{-3}~\mbox{eV}^2$ in normal hierarchy, or $\Delta m^2_{32}$=$-(2.51\pm0.07)\times10^{-3}~\mbox{eV}^2$ and $\Delta m^2_{31}$=$-(2.44\pm0.07)\times10^{-3}~\mbox{eV}^2$ in inverted hierarchy. Both normal and inverted hierarchy are consistent with current experiments. The p-value for normal hierarchy and that for inverted hierarchy are 62\% and 55\%, respectively. This reveals a slight favor for the normal hierarchy, and this preference for normal hierarchy is not disturbed by different {\it CP} phase setting~(Fig.~\ref{figprob}). To draw a stronger conclusion on neutrino mass hierarchy, more accurate measurements on the mass splitting are necessary.

\begin{acknowledgments}
We are grateful to Xinyi Zhang for useful discussions and remarks.
This work is supported by the Principal Fund for Undergraduate Research at Peking
University. It is also partially supported by the National Natural
Science Foundation of China (Grants No.~11035003 and No.~11120101004), and by the National Fund for Fostering
Talents of Basic Science (Grant Nos.~J1103205 and J1103206).
\end{acknowledgments}

\end{document}